\documentclass[aps,prl,reprint,superscriptaddress,amsfonts,amssymb,amsmath,10pt]{revtex4-1}
\usepackage{graphicx}
\usepackage{hyperref}
\usepackage{amsmath}
\usepackage{bm}
\renewcommand{\vec}[1]{\bm{{\mathrm{#1}}}}

\begin{document}

\title{Magnon-photon coupling in a non-collinear magnetic insulator Cu$_2$OSeO$_3$}

\author{L.V. Abdurakhimov}
\email{leonid.abdurakhimov.nz@hco.ntt.co.jp}
\altaffiliation[Present address: ]{NTT Basic Research Laboratories, NTT Corporation, Atsugi, Kanagawa 243-0198, Japan}
\affiliation{London Centre for Nanotechnology, University College London, London WC1H 0AH, United Kingdom}

\author{S. Khan}
\affiliation{London Centre for Nanotechnology, University College London, London WC1H 0AH, United Kingdom}

\author{N.A. Panjwani}
\affiliation{London Centre for Nanotechnology, University College London, London WC1H 0AH, United Kingdom}

\author{J.D. Breeze}
\affiliation{Department of Materials, Imperial College London, Exhibition Road, London SW7 2AZ, United Kingdom}
\affiliation{London Centre for Nanotechnology, Imperial College London, Exhibition Road, London SW7 2AZ, United Kingdom}

\author{M. Mochizuki}
\affiliation{Department of Applied Physics, Waseda University, Okubo, Shinjuku-ku, Tokyo 169-8555, Japan}

\author{S. Seki}
\affiliation{RIKEN Center for Emergent Matter Science (CEMS), Wako 351-0198, Japan}

\author{Y. Tokura}
\affiliation{RIKEN Center for Emergent Matter Science (CEMS), Wako 351-0198, Japan}
\affiliation{Department of Applied Physics and Quantum Phase Electronics Center (QPEC), University of Tokyo, Tokyo 113-8656, Japan}

\author{J.J.L. Morton}
\affiliation{London Centre for Nanotechnology, University College London, London WC1H 0AH, United Kingdom}

\author{H. Kurebayashi}
\email{h.kurebayashi@ucl.ac.uk}
\affiliation{London Centre for Nanotechnology, University College London, London WC1H 0AH, United Kingdom}

\date{\today}

\begin{abstract}
Anticrossing behavior between magnons in a non-collinear chiral magnet Cu$_2$OSeO$_3$ and a two-mode X-band microwave resonator was studied in the temperature range 5--100\,K. In the field-induced ferrimagnetic phase, we observed a strong coupling regime between magnons and two microwave cavity modes with a cooperativity reaching 3600. In the conical phase, cavity modes are dispersively coupled to a fundamental helimagnon mode, and we demonstrate that the magnetic phase diagram of Cu$_2$OSeO$_3$ can be reconstructed from the measurements of the cavity resonance frequency. In the helical phase, a hybridized state of a higher-order helimagnon mode and a cavity mode --- a helimagnon polariton --- was found. Our results reveal a new class of magnetic systems where strong coupling of microwave photons to non-trivial spin textures can be observed.
\end{abstract}

\maketitle

\paragraph{Introduction.}
Strong coupling between microwave photons and particle ensembles is a general phenomenon in light-matter interactions that has been observed in a broad range of condensed-matter systems, including ensembles of magnetically ordered spins\,\cite{Huebl2013,Tabuchi2014,Zhang2014,Goryachev2014,Abdurakhimov2015,Goryachev2018}, paramagnetic spins\,\cite{Chiorescu2010,Schuster2010,Kubo2010,Breeze2017}, and two-dimensional electron systems\,\cite{Muravev2011,Scalari2012,Abdurakhimov2016}. A common feature of ensemble coupling is that the coupling strength between a photon and $N$ particles scales with the square root of $N$, $g_N=g_0\sqrt{N}$, in accordance with the Dicke model\,\cite{Agarwal1984,Immamoglu2009,Garraway2011}. Studies on strong coupling in spin systems are particularly interesting due to possible applications of hybrid spin-ensemble-photon systems for quantum information processing as quantum memories\,\cite{Schoelkopf2008,Kurizki2015} and quantum transducers\,\cite{Blum2015}. The spin-ensemble coupling strength can be extremely large in magnetically ordered systems due to their high spin densities, and extensive studies of strong coupling to magnons -- the quanta of spin wave excitations in magnetically ordered systems --  have been performed recently in experiments on ferrimagnetic insulators. In particular, new magnon-cavity-coupling phenomena have been observed in yttrium iron garnet (YIG), such as coherent coupling between a magnon and a superconducting qubit \cite{Tabuchi2015}, microwave-to-optic-light conversion\,\cite{Hisatomi2016, Haigh2016}, cavity-mediated coherent coupling between multiple ferromagnets\,\cite{Zhang2015, Lambert2016}, spin pumping in a coupled magnon-photon system\,\cite{Bai2015}, and other phenomena\,\cite{Yao2017,Morris2017,Zhang2017,Wang2018,Harder2018}.

\begin{figure}[h]
\includegraphics[width=0.5\textwidth]{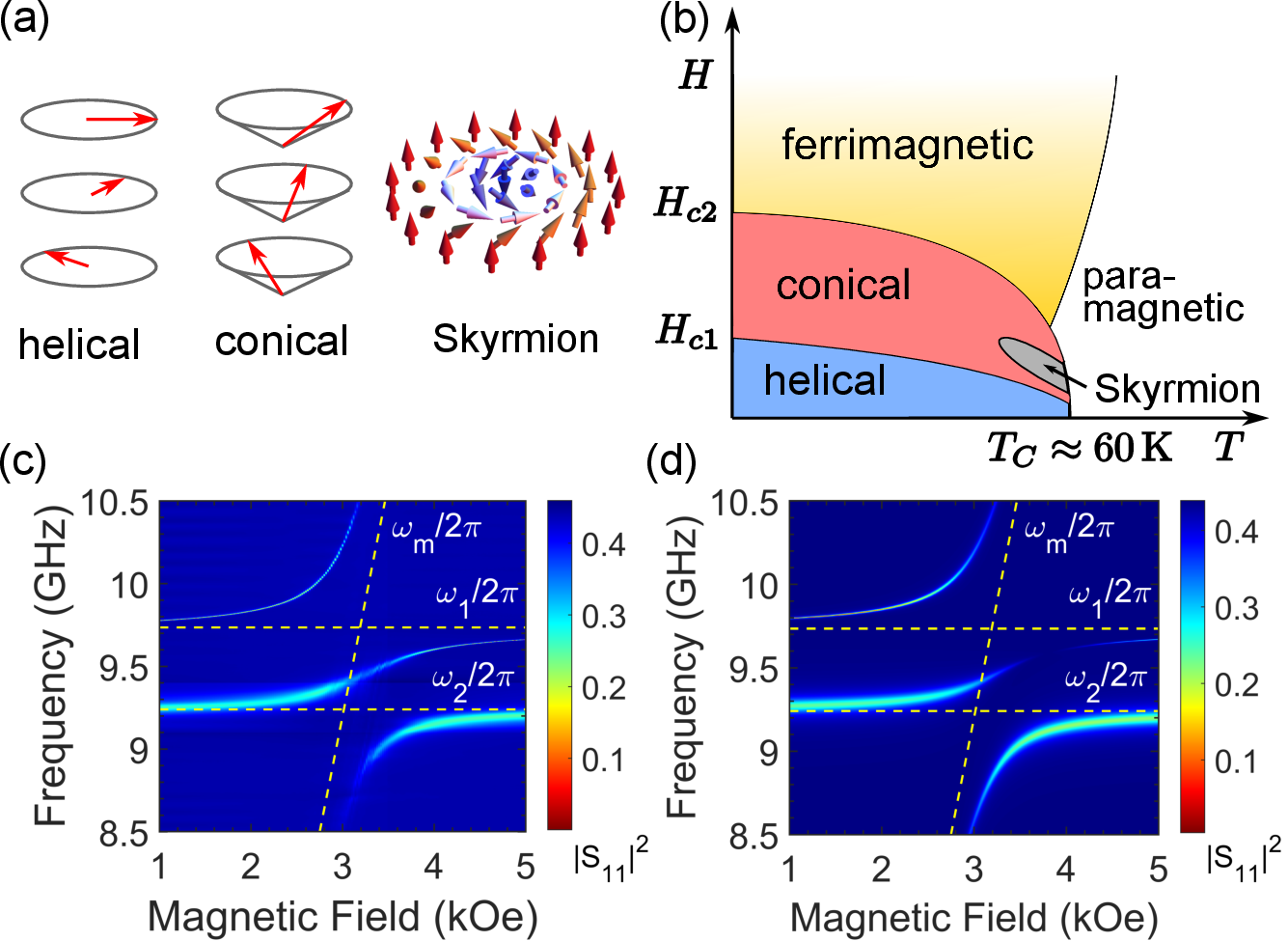}
\caption{\label{figure1}(color online) (a) Schematic view of non-collinear spin textures. (b) Magnetic phase diagram of Cu$_2$OSeO$_3$. (c-d) Strong coupling between ferrimagnetic mode of Cu$_2$OSeO$_3$ and multiple microwave cavity modes: (c) experimental data of microwave reflection $|S_{11}|^2$ as a function of the applied external field and the microwave probe frequency (temperature $T \approx 5$\,K, microwave input power $P \approx 3$\,mW), (d)  microwave reflection $|S_{11}|^2$ calculated using Eq.(\ref{equation-S11}) and parameters described in the text. }
\end{figure}

So far, most studies of strong coupling in magnetic materials have focused on ferrimagnetic materials with the $\vec{S}_i \cdot \vec{S}_j $-like Heisenberg exchange interaction between neighbor spins $\vec{S}_i$ and $\vec{S}_j$. In these materials, all spins are collinear in the ground state, and mainly the uniform precession ferromagnetic mode, or the Kittel mode, has been used in the studies of magnon-photon coupling in those systems. However, there is a growing interest in the coupling of photons to non-collinear and other non-trivial spin systems\,\cite{Osada2018,Mergenthaler2017}. In chiral magnets, the spin-spin exchange interaction consists of two terms; besides the symmetric Heisenberg interaction which favors collinear spin structures, there is an additional antisymmetric $\vec{S}_i \times \vec{S}_j$-like Dzyaloshinskii-Moriya (DM) interaction which tends to twist neighbor spins. As a result of the interplay between Heisenberg and DM exchange interactions, various non-collinear spin textures can be formed in chiral magnets, such as helical, conical, and Skyrmion spin structures (see Fig.\,\ref{figure1}(a)). Studies of the coupling between microwave photons and non-trivial spin textures is a potentially rich and largely unexplored area.

A chiral magnetic insulator copper-oxoselenite Cu$_2$OSeO$_3$ crystallizes in a non-centrosymmetric cubic structure with 16 copper ions Cu$^{2+}$ per unit cell (space group $P2_13$, lattice constant $a\approx8.93${\AA}\,\cite{Larranga2009}). The basic magnetic building block of Cu$_2$OSeO$_3$ is a tetrahedral cluster formed by four Cu$^{2+}$ spins in a 3-up-1-down spin configuration, which behaves as a spin triplet with the total spin $S=1$\,\cite{Janson2014}. This magnetic structure has been visualized elsewhere\,\cite{Seki2012a,Janson2014}. Due to a combination of Heisenberg and DM exchange interactions, the system of $S=1$ clusters forms helical, conical, ferromagnetic and Skyrmion magnetic phases  in an applied external magnetic field below the Curie temperature $T_{\rm C}\approx 60$\,K, as shown in Fig.\,\ref{figure1}(b). Above $T_{\rm C}$, the system is paramagnetic. Besides being a chiral magnet, Cu$_2$OSeO$_3$ is also of particular interest due to its multiferroic and magnetoelectric properties\,\cite{Seki2012a,Seki2012b,Okamura2013,Mochizuki2015,Ruff2015}.

In this Letter, we report a study of magnon-photon coupling in a Cu$_2$OSeO$_3$ system. In the collinear ferrimagnetic phase, we observed a strong coupling regime between a microwave cavity mode and a uniform Kittel magnon mode, and the temperature dependence of the coupling strength was found to follow that of the square root of the net magnetization. In the non-collinear conical phase, a dispersive coupling regime between helimagnons and microwave photons was observed, and we demonstrate that the magnetic phase diagram can be determined by the measurement of the frequency of a cavity mode. In the non-collinear helical phase, normal-mode splitting was detected between a higher-order helimagnon mode and a cavity mode.  

\paragraph{Experimental details.}
We performed microwave X-band spectroscopy studies of Cu$_2$OSeO$_3$ in the temperature range 5--100\,K using a helium-flow cryostat\,\cite{SM}. A sample of single crystal Cu$_2$OSeO$_3$ was inserted into a commercial Bruker MD5 microwave cavity consisting of a sapphire dielectric ring resonator mounted inside a metallized plastic enclosure. The sample mass was $m \approx 60$\,mg, corresponding to a total effective spin number $N \approx 7 \times 10^{19}$. The shape of the sample was close to semi-ellipsoidal, with the lengths of semi-axes being 1.5\,mm, 1.5\,mm, and 2\,mm, and a flat plane being oriented along the long ellipsoid axis.  The orientation of crystallographic axes of the sample relative to the cavity axis was chosen arbitrarily. The cavity supported two microwave modes\,\cite{SM,Mongia1994}: the primary mode TE$_{01\delta}$ with the resonance frequency of about $\omega_1/2 \pi \approx$\,9.74\,GHz and the hybrid mode HE$_{11\delta}$ with the resonance frequency of about $\omega_2/2 \pi \approx$\,9.24\,GHz. We tuned the quality factor $Q_1$ of the primary-mode resonance by adjusting the position of a coupling loop antenna. In our measurements, we used a slightly under-coupled cavity with $Q_1 \approx 5 \times 10^3$. The quality factor of the hybrid-mode resonance did not depend on the position of the coupling antenna, and was about $Q_2 \approx 100$. In our experiments, the microwave reflection S-parameter $|S_{11}|^2$ was measured as a function of external magnetic field and microwave probe frequency.

\paragraph{Results.}
Figure\,\ref{figure1}(c) shows typical data from microwave reflection measurements at  temperature $T \approx 5$\,K, obtained from raw experimental data by background-correction processing\,\cite{SM}. The input microwave power was $P \approx 3$\,mW. Two avoided crossings are visible at the degeneracy points where two cavity modes would otherwise intersect a magnon mode. The magnon mode corresponds to a uniform spin precession (Kittel mode) with frequency $\omega_m/2\pi = \gamma (H_0 + H_{\rm demag})$, where $H_0$ is the applied magnetic field, $H_{\rm demag}$ is the demagnetizing field, and $\gamma \approx 28$\,GHz/T is the electron gyromagnetic ratio. Here, we assume that anisotropy fields are small and can be neglected.

\begin{figure*}
\includegraphics[width=0.9\textwidth]{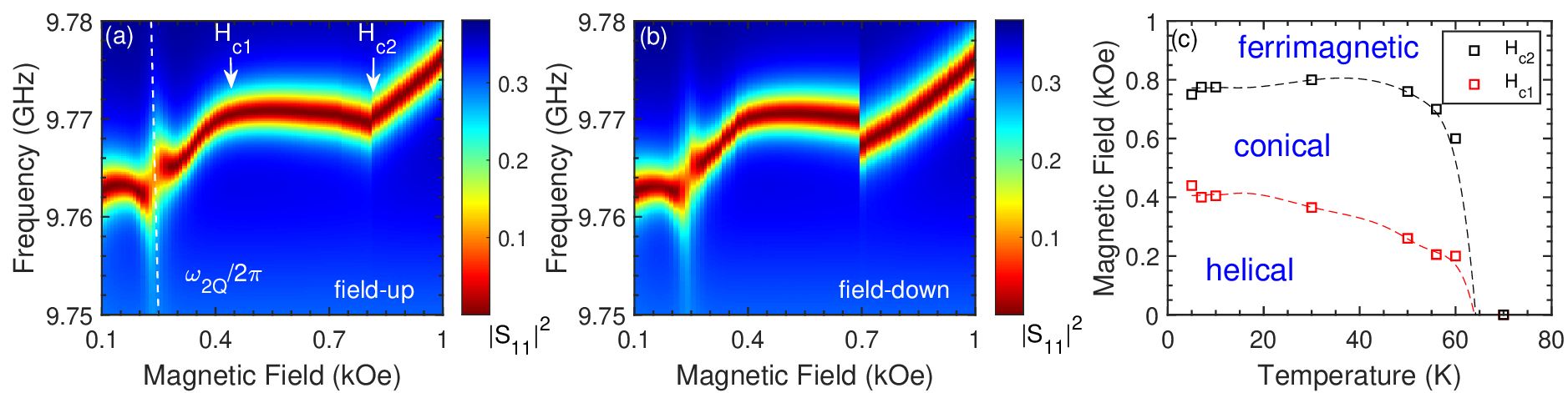}
\caption{\label{figure-low-field-features}(color online) Magnon-photon coupling in non-collinear magnetic phases. (a) Microwave reflection $|S_{11}|^2$ at 5\,K. The magnetic field sweep was performed from low to high field values (``field-up''). Dashed line corresponds to a higher-order helimagnon $k=\pm 2Q$ mode, and an avoided crossing between the helimagnon mode and the cavity mode is clearly visible. $H_{c1}$ and $H_{c2}$ are critical magnetic fields of helical-to-conical and conical-to-ferrimagnetic phase transitions, respectively. (b) Microwave reflection $|S_{11}|^2$ at 5\,K. The magnetic field sweep was performed from high to low values (``field-down''). The transition from ferrimagnetic to conical phase demonstrated hysteretic behavior. (c) The magnetic phase of Cu$_2$OSeO$_3$ sample reconstructed from the ``field-up'' measurements of $H_{c1}$ and $H_{c2}$. The dashed lines are for eye guidance.}
\end{figure*} 

The interaction between two cavity modes and a magnon mode can be described by the following Hamiltonian in the rotating-wave approximation (RWA):
\small
\begin{align}
\mathcal{H}_0 / \hbar & = \omega_1 a_1^{\dagger} a_1 +\omega_2 a_2^{\dagger} a_2 + \omega_m m^{\dagger} m + \nonumber \\
& + g_1 (a_1^{\dagger}m+a_1 m^{\dagger}) + g_2 (a_2^{\dagger}m+a_2 m^{\dagger}),
\label{equation-hamiltonian}
\end{align}
\normalsize
where $a_1^{\dagger}$ ($a_1$) is the creation (annihilation) operator for microwave photons at frequency $\omega_1$, $a_2^{\dagger}$ ($a_2$) is the creation (annihilation) operator for microwave photons at frequency $\omega_2$, $m^{\dagger}$ ($m$) is the creation (annihilation) operator for magnons at frequency $\omega_m$, and $g_1$ ($g_2$) is the coupling strength between the magnon mode and the first (second) cavity mode.

In order to extract numerical values of coupling strengths $g_1$ and $g_2$ and other parameters from the experimental data, we used the following equation obtained from input-output formalism theory\,\cite{SM}:
\small
\begin{equation}
|S_{11}|^2= \left| -1+\frac{\kappa_1^{\mathrm{(c)}}F_2 + \kappa_2^{\mathrm{(c)}}F_1 - 2 \sqrt{\kappa_1^{\mathrm{(c)}} \kappa_2^{\mathrm{(c)}}}F_3}{F_1 F_2 - F_3^2} \right| ^2,
\label{equation-S11}
\end{equation}
\normalsize
where 
\small
\begin{align*}
F_1 &= i(\omega_1-\omega)+\left(\kappa_1+\kappa_1^{\mathrm{(c)}}\right)/2 + g_1^2(i(\omega_m-\omega)+\gamma_m/2)^{-1}, \\
F_2 &= i(\omega_2-\omega)+\left(\kappa_2+\kappa_2^{\mathrm{(c)}}\right)/2 + g_2^2(i(\omega_m-\omega)+\gamma_m/2)^{-1}, \\
F_3 &= \sqrt{\kappa_1^{\mathrm{(c)}} \kappa_2^{\mathrm{(c)}}}/2 + g_1 g_2 (i(\omega_m-\omega)+\gamma_m/2)^{-1},
\end{align*}
\normalsize
and $\kappa_1$ ($\kappa_2$) is the damping rate of the first (second) cavity mode, $\kappa_1^{\mathrm{(c)}}$ ($\kappa_2^{\mathrm{(c)}}$) is the coupling rate between the first (second) cavity mode and the output transmission line, and $\gamma_m$ is the damping rate of the magnonic mode. Damping rates represent linewidths (FWHM) of the corresponding modes. 

We reproduce the data shown in Fig.\,\ref{figure1}(c) by using Eq.\,(\ref{equation-S11}) with the following parameters: $g_1/2\pi \approx 600$\,MHz, $g_2/2\pi \approx 450$\,MHz, $\kappa_1/2\pi \approx 1$\,MHz,  $\kappa_1^{\mathrm{(c)}}/2\pi\approx 1$\,MHz, $\kappa_2/2\pi \approx 60$\,MHz,  $\kappa_2^{\mathrm{(c)}}/2\pi\approx 10$\,MHz, $H_{demag}\approx 280$\,Oe and $\gamma_m/2\pi \approx 50$\,MHz (see Fig.\,\ref{figure1}(d)). Thus, the coupling strengths are much greater than the damping rates of both cavity and magnon modes, $g_1 \gg (\kappa_1+\kappa_1^{\mathrm{(c)}}), \gamma_m$ and $g_2 \gg (\kappa_2+\kappa_2^{\mathrm{(c)}}), \gamma_m$, and strong coupling regimes are realized for both avoided crossings. From the obtained value of the ferrimagnetic resonance linewidth $\gamma_m$, we estimate the Gilbert damping parameter $\alpha \approx 2.6 \times 10^{-3}$ at 5\,K which is consistent with the literature\,\cite{Stasinopoulos2017,damping}. Cooperativity parameters are much higher than unity: $C_1 = g_1^2/\left(\gamma_m \left(\kappa_1+\kappa_1^{\mathrm{(c)}}\right)  \right) \approx 3600$ and $C_2 = g_2^2/\left(  \gamma_m \left(\kappa_2+\kappa_2^{\mathrm{(c)}}\right) \right) \approx 60$. Moreover, ratios $g_1/\omega_1 \approx 0.06$ and $g_2/\omega_2 \approx 0.05$ are close to the condition of the ultra-strong coupling regime ($g/\omega \geq 0.1$), where the coupling strength is comparable with the frequency of the degeneracy point of an avoided crossing, and new physics beyond RWA can be explored\,\cite{Zhang2014,Goryachev2014}.

The obtained values of coupling strengths $g_1$ and $g_2$ between magnons and cavity modes are in relatively good agreement with the theoretical estimates
\begin{equation}
g_i^\mathrm{(th)} = \frac{\eta_i}{2} \gamma \sqrt{\frac{\mu_0 \hbar \omega_i}{V_i}} \sqrt{2SN}, \quad (i = 1,2)
\label{equation-g}
\end{equation}
where $\mu_0$ is the vacuum permeability, $V_1 \approx 230$\,mm$^3$ ($V_2 \approx 290$\,mm$^3$) is the mode volume of the primary (hybrid) cavity resonance, and  the coefficient $\eta_1 \approx 0.85$ ($\eta_2 \approx 0.83$) describes the spatial overlap between the primary (hybrid) cavity mode and the magnon mode\,\cite{SM}. Substituting other known parameters into the equations, we obtain $g_1^\mathrm{(th)}/2\pi \approx 825$\,MHz and $g_2^\mathrm{(th)}/2\pi \approx 700$\,MHz. Slight discrepancies between theoretical and experimental values of coupling strengths can be caused by the excitation of additional $k \neq 0$ spin-wave modes in the sample which are visible as additional faint narrow lines in the experimental data, and the resulting reduction of the effective number of spins involved in the uniform Kittel mode precession\,\cite{Boventer2018}. The difference in coupling strengths can be also related to the fact that the coupled system of magnons and photons is close to the ultra-strong coupling regime mentioned above where the Hamiltonian (\ref{equation-hamiltonian}) is not valid.

Coupling strengths $g_1$ and $g_2$ depended strongly on temperature\,\cite{SM}. In the ferrimagnetic phase, the effective number of spins is proportional to the net magnetization, and we found that the temperature dependence of the coupling rate could be fitted by square-root function of the net magnetization which is in good agreement with the results of studies of strong coupling in collinear ferrimagnetic systems\,\cite{Maier-Flaig2017}. In our experiments, anticrossing behavior was independent of microwave probe power, consistent with observations of strong coupling in other systems\,\cite{Chiorescu2010,Huebl2013,Abdurakhimov2015}.

\begin{figure}[t]
\includegraphics[width=0.5\textwidth]{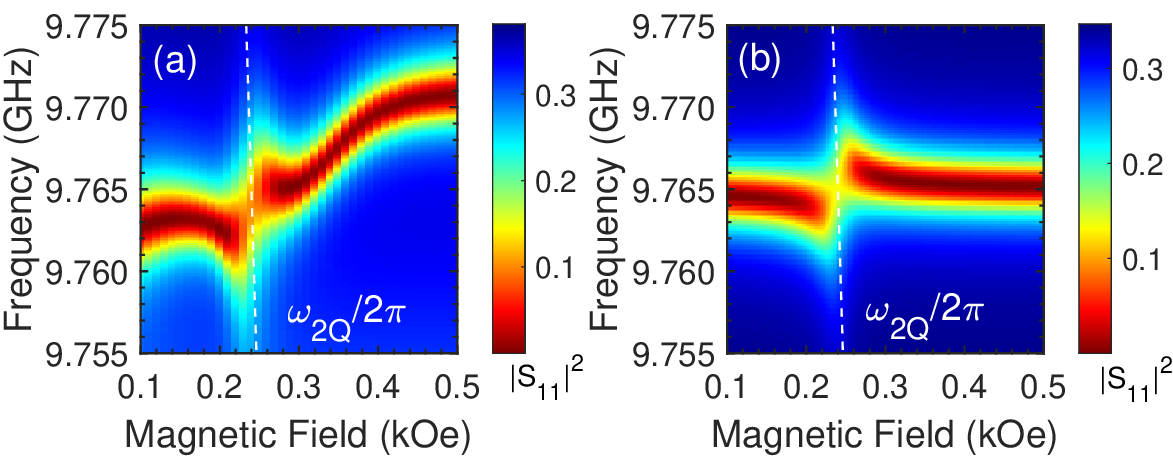}
\caption{\label{figure-helimagnon-coupling}(color online) Normal-mode splitting between a cavity mode and a higher-order helimagnon mode in the helical phase. (a) Microwave reflection $|S_{11}|^2$ at 5\,K. (b) Results of the numerical calculation of $|S_{11}|^2$ by using equations and parameters described in the text.}
\end{figure}

In order to study magnon-photon coupling in non-collinear spin textures, we performed measurements at low magnetic fields where the system exhibits helical and conical magnetic phases (see Fig.\,\ref{figure-low-field-features},\ref{figure-helimagnon-coupling}). We found that the frequency of a cavity resonance depended on the applied external magnetic field not only in the ferrimagnetic phase, but also in the helical and conical states (Fig.\,\ref{figure-low-field-features}(a-b)). We identify two features in microwave response at magnetic field values $H_{c1}$ and $H_{c2}$  which can be attributed to helical and conical magnetic phase transitions in Cu$_2$OSeO$_3$, respectively (see Fig.\,\ref{figure-low-field-features}(c)). The observed transition from helical to conical phases is relatively smooth, which can be related to the fact that in the helical phase the spin system forms a multidomain structure of flat helices\,\cite{Seki2012a}, and, since the external dc magnetic field was not aligned along the high symmetry directions of the crystal structure in our measurements, domains gradually reoriented themselves with the increase of the magnetic field\,\cite{Schwarze2015}.

We suppose that the changes in the frequencies of cavity modes are caused by dispersive coupling between microwave photons and magnonic modes in the helical and conical phases --- so-called helimagnons\,\cite{Schwarze2015,Weiler2017}. Frequencies of the fundamental $n=\pm 1$ helimagnon modes $\omega_{\pm Q}$ lie well below the cavity resonance frequencies and depend weakly on the applied magnetic field. Since a full theoretical model of helimagnon-photon coupling would require detailed calculation of spectral weights of helimagnon modes\,\cite{Schwarze2015} which is outside the scope of this paper, here we present a qualitative description of the dispersive magnon-photon coupling in Cu$_2$OSeO$_3$. In the helical phase $H<H_{c1}$, the net magnetization $M$ is small\,\cite{Seki2012a}, and helimagnon-photon coupling is very weak. In the conical phase $H_{c1} < H < H_{c2}$, the net magnetization is substantial, and the cavity mode is dispersively coupled to the fundamental helimagnon mode which results in the shift of the cavity resonance frequency.  In the ferrimagnetic phase $H>H_{c2}$, the net magnetization is close to its maximum value (saturation magnetization), the coupling strength to the Kittel mode is large, and the shift of the cavity resonance with the increase of the magnetic field is large. It should be noted that the magnetic-field dependence of the cavity resonance frequency cannot be explained by the variation of a dc magnetic permeability of the material\,\cite{SM}.

The value of the magnetic field $H_{c2}$ of the conical-to-ferrimagnetic transition  was found to be dependent on the direction of magnetic field sweep (Fig.\,\ref{figure-low-field-features}(a-b)). This hysteretic behavior was observed only at low temperatures $T\lesssim 40$\,K, and it can be related to either the competition between the conical state and the unusual ``tilted conical'' and skyrmion states recently reported in\,\cite{Qian2018,Chacon2018} or the extension of the conical $n=1$ mode into the induced-ferrimagnetic phase (and vice versa)\,\cite{Weiler2017}.  

In the helical magnetic phase, we observed hybridization between a cavity mode and a higher-order helimagnon mode (see Fig.\,\ref{figure-helimagnon-coupling}). By analogy with cavity magnon polaritons\,\cite{Bai2015,Goryachev2018,Yao2017,Zhang2017}, a hybrid helimagnon-photon state can be called a helimagnon polariton. In contrast to the ferrimagnetic mode, the dispersion curve of a helimagnon mode $\omega_{nQ}(H)$ exhibits a negative slope ($d\omega_{nQ}/dH < 0$). The helical phase is characterized by a multidomain structure of flat helices, where the propagation vectors of the different helices are pinned along the preferred axes of the system.  The accurate description of helimagnons in the helical phase requires taking into account the cubic anisotropies of Cu$_2$OSeO$_3$\,\cite{Schwarze2015}, which is outside the scope of this paper. Instead, we can use the analyitcal equation used for the description of magnons in the conical phase\,\cite{Weiler2017}:
\begin{equation}
\omega_{nQ} = |n| \frac{\gamma B_{c2}}{1+N_d \cdot \chi} \sqrt{n^2 + (1+\chi)(1-(B_0/B_{c2})^2)},
\end{equation}
where $N_d$ is the demagnetization factor along the direction of the $Q$ vector, $\chi$ is the internal conical susceptibiliy ($\chi \approx 1.76$ for Cu$_2$OSeO$_3$\,\cite{Schwarze2015}), $B_0$ is the applied external field, and $B_{c2}$ is the critical field for the transition between conical and ferrimagnetic phases ($B_{c2}\approx 0.08$\,T in our experiments). The mode number $n$ describes the relation between the helimagnon wave vector $k$ and the wave vector of the helical spiral $Q$, $k = \pm nQ$. By adjusting the parameters $n$ and $N_d$, we identify the observed helimagnon mode as an $n=\pm 2$ mode (the demagnetization factor $N_d \approx 0.1$), which is shown by the dashed line in Fig.\,\ref{figure-helimagnon-coupling}.

In order to characterize the observed avoided crossing quantitatively, we use the following equation for the microwave reflection\,\cite{Zhang2014}:
\begin{equation}
|S_{11}|^2 = \left|-1 + \frac{\kappa_1^{\mathrm{(c)}}}{i(\omega_1 - \omega) + \frac{\kappa_1 + \kappa_1^{\mathrm{(c)}}}{2} + \frac{g_{2Q}^2}{i(\omega_{2Q}-\omega)+\frac{\gamma_{2Q}}{2}}} \right|^2,
\end{equation}
where $\omega_{2Q}$ and $\gamma_{2Q}$ are the frequency and the linewidth of the helimagnon mode, respectively, and $g_{2Q}$ is the coupling strength between the cavity mode and the helimagnon mode. We found that experimental data shown in Fig.\,\ref{figure-helimagnon-coupling}(a) can be reproduced by using the following parameters: $g_{2Q}/2\pi \approx 8.5$\,MHz, $\kappa_1/2\pi \approx 1.5$\,MHz, $\kappa_1^{\mathrm{(c)}}/2\pi \approx 1.5$\,MHz, and $\gamma_{2Q}/2\pi \approx 60$\,MHz (see Fig.\,\ref{figure-helimagnon-coupling}(b)). Therefore, the observed normal-mode splitting can be attributed to the Purcell effect ($\kappa < g_{2Q} < \gamma_m$), where the decay of microwave cavity photons is enhanced due to their interaction with lossy magnons\,\cite{Zhang2014}. We could not detect the higher-order helimagnon mode $\omega_{2Q}$ at temperatures $T \gtrsim 30$\,K which is consistent with the literature\,\cite{Weiler2017}.

According to our numerical simulations, the mode overlapping between a cavity mode and higher-order $|n|=2$ helimagnon modes is suppressed as compared to the one for the Kittel ferromagnetic mode\,\cite{SM}. It should be noted that, according to the work\,\cite{Weiler2017}, the direct interaction between a uniform microwave mode and $|n|=2$ helimagnons is negligible, but microwave photons can couple to $|n|=2$ helimagnons indirectly via fundamental $|n|=1$ helimagnons. Indeed, in the coordinate frame co-rotating with the spins around the helical-spiral wave vector $\vec{Q} \parallel \vec{z}$, the magnetic component of the TE$_{01\delta}$ cavity mode --- which is spatially uniform within the sample volume in the laboratory frame of coordinates --- corresponds to an effective microwave field with components $h_{1x}^{eff} \propto \cos{(Qz)}$, and $h_{1y}^{eff} \propto \sin{(Qz)}$. Therefore, the applied microwave magnetic field can excite directly only $|n|=1$ helimagnons which are characterized by the wave vector $|k|=Q$. However, in a cubic crystal, due to the fourth-order magnetic anisotropy $m_x^4+m_y^4+m_z^4$, where $\vec{m}$ is a unit vector in the direction of the magnetization, $n=\mp1$ and $n=\pm2$ helimagnon modes are hybridized\,\cite{Weiler2017}. Thus, the observed avoided crossing is caused by the double hybridization between microwave photons, $|n|=1$ helimagnons, and $|n|=2$ helimagnons. 

We were not able to resolve coupling to magnetic excitations in the Skyrmion phase in the measurements in the corresponding temperature range, presumably due to large detuning between the microwave cavity mode and Skyrmion modes\,\cite{skyrmion}. 

\paragraph{Conclusions.}
We performed a study of magnon-photon coupling in helical, conical and ferrimagnetic phases of a chiral magnet Cu$_2$OSeO$_3$. We achieved a strong coupling regime between a ferrimagnetic magnon mode and multiple microwave cavity modes. In the non-collinear conical phase, we observed the dispersive coupling between cavity modes and a fundamental helimagnon mode which allowed us to use a cavity mode as a probe for the sensing of magnetic phase transitions in Cu$_2$OSeO$_3$. In the non-collinear helical phase, we detected a normal-mode splitting between microwave photons and high-order helimagnons in the Purcell-effect regime. These findings establish a new area of studies of strong coupling phenomena in multiferroic chiral magnetic systems, paving the way for new hybrid systems consisting of non-trivial spin textures coupled to microwave photons via magnetic and magnetoelectric interactions.

\begin{acknowledgments}
This work was partly supported by the Grants-In-Aid for Scientific Research (Grants Nos. 18H03685, 17H05186 and 16K13842) from JSPS, and by the European Union's Horizon 2020 programme (Grant Agreement Nos. 688539 (MOS-QUITO) and 279781 (ASCENT)); as well as the Engineering and Physical Science Research Council UK through UNDEDD project (EP/K025945/1). 
\end{acknowledgments}

%

\widetext
\clearpage
\setcounter{equation}{0}
\setcounter{figure}{0}
\setcounter{table}{0}
\setcounter{page}{1}
\makeatletter
\renewcommand{\theequation}{S1-\arabic{equation}}
\renewcommand{\thefigure}{S1-\arabic{figure}}
\renewcommand{\bibnumfmt}[1]{[S1-#1]}
\renewcommand{\citenumfont}[1]{S1-#1}

\begin{center}
\textbf{\large Supplemental Material for ``Magnon-photon coupling in a non-collinear magnetic insulator Cu$_2$OSeO$_3$''. Part 1.}
\end{center}

\section{Experimental details}

\begin{figure}[h]
\includegraphics[width=0.3\textwidth]{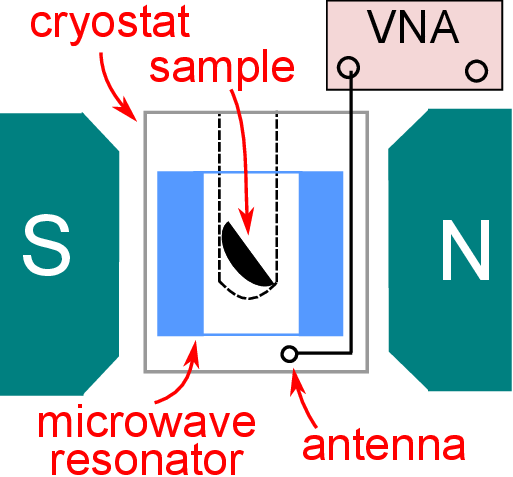}
\caption{\label{figure-experimental-setup}(color online) Schematic view of the experimental setup.}
\end{figure}

Experiments were performed using the experimental setup described in the main text and schematically shown in Figure\,\ref{figure-experimental-setup}.

\begin{figure}[h]
\includegraphics[width=0.8\textwidth]{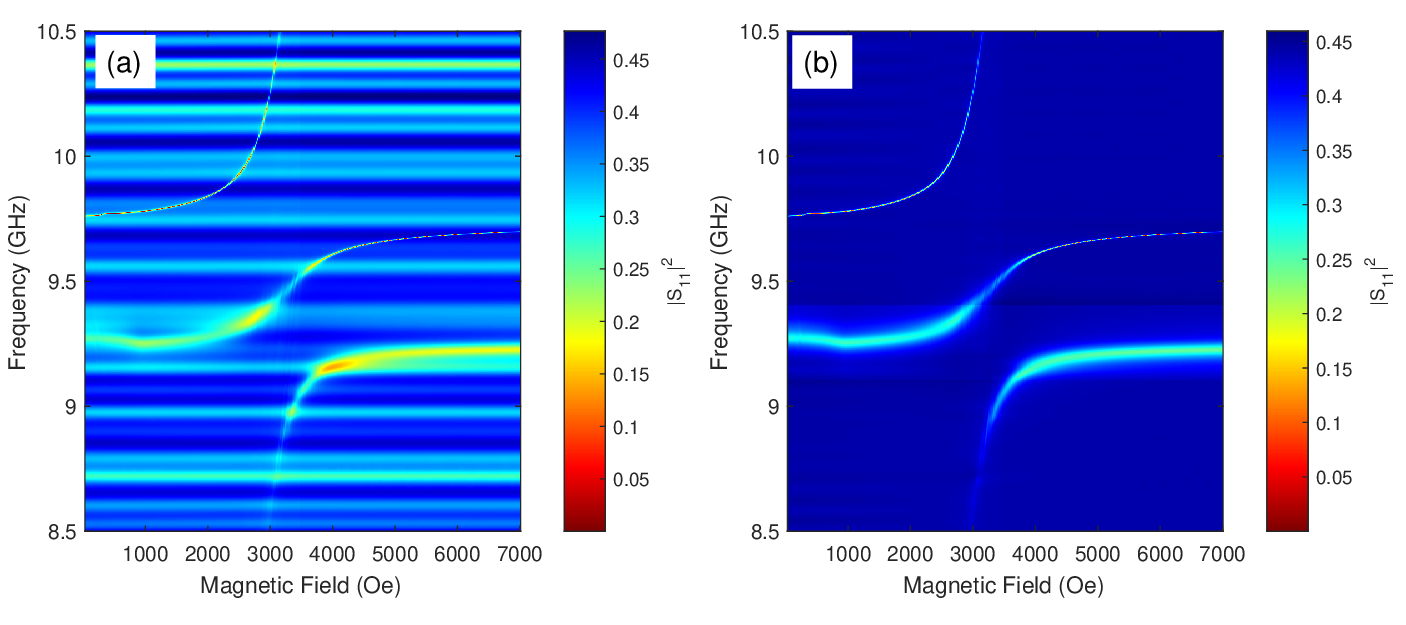}
\caption{\label{figure-background-correction}(color online) The standing-wave background correction. (a) An example of raw experimental data ($T=5$\,K). (b) The same data after the correction.}
\end{figure}

An example of raw experimental data is shown in Figure\,\ref{figure-background-correction}(a). We can clearly see two avoided crossings and a background pattern consisting of horizontal stripes. The background is caused by the presence of a standing wave in coaxial cables between the microwave resonator and the vector network analyzer due to impedance mismatch at the input of the microwave resonator. To eliminate that background, we reconstructed the standing-wave profile by a piecewise-defined function with different pieces being taken at different values of magnetic field, where the standing-wave background was not affected by the avoided crossings. Background-corrected data was obtained by subtracting the standing-wave profile from the raw experimental data, and adding an offset value to keep the minimum and maximum levels of $|S_{11}|^2$ at the original level. An example of background-corrected data is shown in Figure\,\ref{figure-background-correction}(b).

\begin{figure}[h]
\includegraphics[width=0.8\textwidth]{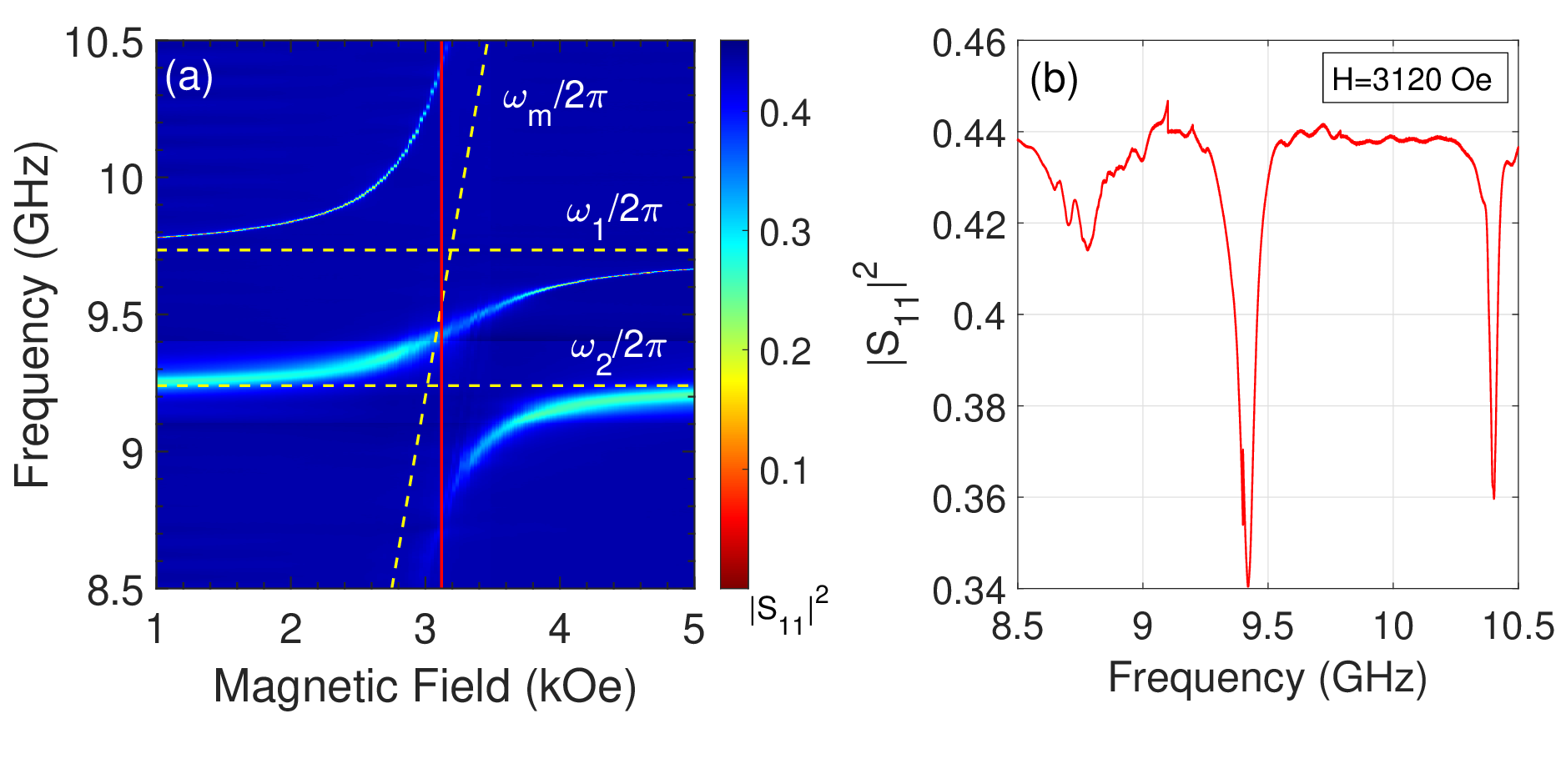}
\caption{\label{figure-line-scan}(color online) Normal-mode splitting due to strong coupling between a ferrimagnetic mode and two cavity modes (see the main text for other details). (a) Microwave reflection $|S_{11}|^2$ as a function of the applied external field and the microwave probe frequency (temperature $T \approx 5$\,K). (b) The frequency scan at the magnetic field $H_0 = 3120$\,Oe. The splitting of two cavity modes into three hybrid resonance modes is clearly visible. The additional fine structure of the low-frequency resonance is due to excitation of additional standing spin wave resonances.}
\end{figure}

The strong coupling regime between a ferrimagnetic mode and two cavity modes was manifested as splitting of two cavity resonances into three hybrid resonant modes (see Fig.\,\ref{figure-line-scan}).  

\newpage
\section{Microwave TE$_{01\delta}$ and HE$_{11\delta}$ modes of a dielectric resonator}

\begin{figure}[h]
\includegraphics[width=0.9\textwidth]{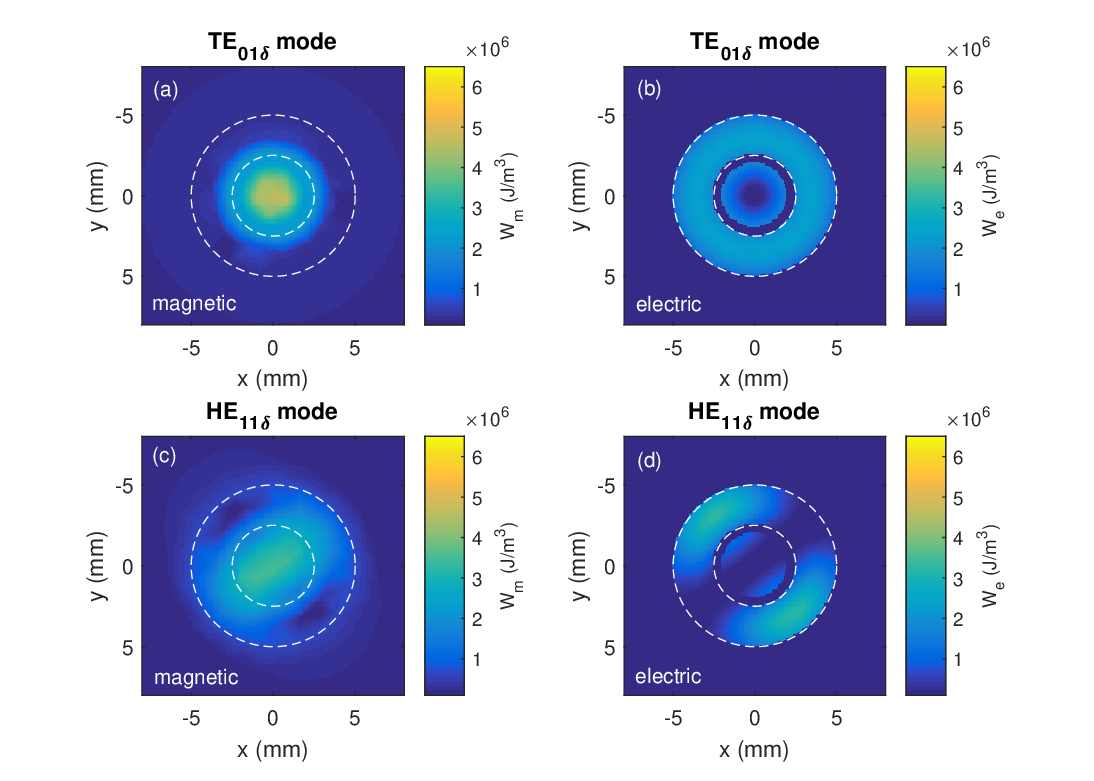}
\caption{\label{figure-hybrid-mode}(color online) Results of numerical simulations of the dielectric resonator cavity. Distributions of magnetic ($W_m$) and electric ($W_e$) energy densities in the center horizontal plane, which is perpendicular to the axis of the dielectric ring resonator, are shown for the primary cavity mode (a-b) and the hybrid cavity mode (c-d). Dashed white circles represent the contour of the dielectric ring resonator.}
\end{figure}

We performed numerical simulations of the sapphire dielectric ring resonator cavity by using CST Microwave Studio software (see Fig.\,\ref{figure-hybrid-mode}). The sample was modeled as a dielectric sphere with the radius $r=2$\,mm, the relative magnetic permeability $\mu_r = 1$, and the relative electrical permittivity $\varepsilon_r \approx 12.5$\,\cite{SM1-Ruff2015}. We found that the cavity supports two microwave modes: a primary TE$_{01\delta}$ mode with the resonance frequency of about $\omega_1/2 \pi \approx$\,9.8\,GHz (Fig.\,\ref{figure-hybrid-mode}(a-b)) and a hybrid HE$_{11\delta}$ mode with the resonance frequency of about $\omega_2/2 \pi \approx$\,9.56\,GHz (Fig.\,\ref{figure-hybrid-mode}(c-d)). The microwave magnetic field of the primary mode is parallel to the axis of the dielectric resonator, and magnetic energy density $W_m$ is concentrated inside the cavity bore (Fig.\ref{figure-hybrid-mode}(a)). The effective magnetic mode volume for the primary cavity resonance can be estimated as
\begin{equation}
V_1 = \frac{\int (\frac{1}{2}\varepsilon_0 \varepsilon_r E^2 + \frac{1}{2}\mu_0 \mu_r H^2)dV}{\mu_0 \mu_r H_{max}^2}=\frac{\int (W_e + W_m) dV}{2W_{m,max}},
\label{equation-V1}
\end{equation}
where $\varepsilon_0$ is the vacuum permittivity, $\varepsilon_r = \varepsilon_r(x,y,z)$ is the relative electrical permittivity at a given point, $E=E(x,y,z)$ is the microwave electric field, $\mu_0$ is the vacuum permeability, $\mu_r=\mu_r(x,y,z)$ is the relative permeability, $H = H(x,y,z)$ is the microwave magnetic field, $H_{max}$ is the maximum value of the microwave magnetic field, $W_e$ and $W_m$ are microwave electrical and magnetic energy densities, and the integration is taken over the total volume of the system. From numerical simulations, we obtain $V_1 \approx 230$\,mm$^3$.

Next, we introduce an overlap coefficient $\eta_1$ to take into account a slight non-uniformity of the distribution of the microwave magnetic field across the sample which affects our calculations of coupling strengths described in the main text:
\begin{equation}
\eta_{1} = \sqrt{\frac{\int_{V_s} \mu_0 \mu H^2 dV}{\mu_0 \mu H_{max}^2 V_s}} = \sqrt{\frac{\int_{V_s} W_m dV}{W_{m,max} V_s}},
\label{equation-eta1}
\end{equation}
where $V_{s} = \frac{4}{3} \pi r^3$ is the volume of the sample used in the numerical model, and the integration is taken over the sample volume. Numerically, we find $\eta_{1} \approx 0.85$ for the primary mode. 

The hybrid HE$_{11\delta}$ mode is a double-degenerate axially-asymmetric mode. At the site of the sample, the magnetic component lies in the plane perpendicular to the axis of the dielectric resonator. The HE$_{11\delta}$ mode is degenerate, and two mutually perpendicular orientations of microwave magnetic field --- two polarizations --- are possible. Typical distributions of microwave magnetic and electric energy densities for one of two possible polarizations are shown in Fig.\,\ref{figure-hybrid-mode}(c-d)). In calculations of the coupling strength described in the main text, we take into account only one polarization of the HE$_{11\delta}$ mode for which the microwave magnetic field at the site of the sample is perpendicular to the external dc magnetic field (a hybrid mode with another polarization does not interact with a ferrimagnetic mode). By using equations similar to Eq.\,(\ref{equation-V1}--\ref{equation-eta1}), we estimate that the effective magnetic mode volume is $V_{2} \approx 290$\,mm$^3$, and the overlap coefficient is $\eta_2 \approx 0.83$ for the magnetic component of the hybrid mode. 

\newpage
\section{Input-output formalism: strong coupling between a magnon mode and a two-mode cavity}

Following input-output formalism theory \cite{Gardiner2000, Walls2008}, we can write the following Heisenberg-–Langevin equations:
\begin{align*}
\dot{a}_1(t) = - \frac{i}{\hbar}[a_1(t),\mathcal{H}_{\mathrm{SYS}}] - \frac{\kappa_1^{\mathrm{(cpl)}}}{2}a_1(t) - \frac{\sqrt{\kappa_1^{\mathrm{(cpl)}} \kappa_2^{\mathrm{(cpl)}}}}{2} a_2(t) +\sqrt{\kappa_1^{\mathrm{(cpl)}}} a^{\mathrm{(in)}}(t), \\
\dot{a}_2(t) = - \frac{i}{\hbar}[a_2(t),\mathcal{H}_{\mathrm{SYS}}] - \frac{\kappa_2^{\mathrm{(cpl)}}}{2}a_2(t)- \frac{\sqrt{\kappa_1^{\mathrm{(cpl)}}\kappa_2^{\mathrm{(cpl)}}}}{2} a_1(t) +\sqrt{\kappa_2^{\mathrm{(cpl)}}} a^{\mathrm{(in)}}(t), \\
\dot{m}(t) = -\frac{i}{\hbar}[m(t),\mathcal{H}_{\mathrm{SYS}}], \\
a_{in}(t)+a_{out}(t) = \sqrt{\kappa_1^{\mathrm{(cpl)}}} a_1(t)+ \sqrt{\kappa_2^{\mathrm{(cpl)}}} a_2(t),
\end{align*}
where $a_{in}$ and $a_{out}$ are an external input and output fields, $\kappa_1^{\mathrm{(cpl)}}$ and $\kappa_1^{\mathrm{(cpl)}}$ are coupling constants between cavity fields and the external fields. The Hamiltonian $\mathcal{H}_{\mathrm{SYS}}=\mathcal{H}_0+\mathcal{H}_{\mathrm{B}}$ describes the cavity modes and the magnon mode, where the Hamiltonian $\mathcal{H}_0$ is defined in the main text, and $\mathcal{H}_{\mathrm{B}}$ describes the dissipation inside the cavity due to the interaction with a heat bath. 

\begin{figure}[h]
\includegraphics{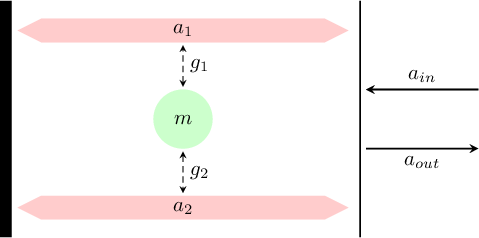}
\caption{\label{figure-input-output-formalism}(color online) A schematic representation of the two cavity fields, the magnon mode, and the input and output fields for a single-sided cavity.}
\end{figure}

By assuming that solutions should be in the form of $a_1(t)= a_1 e^{-i \omega t}$, $a_2(t)= a_2 e^{-i \omega t}$, etc., we can obtain the following equations:
\begin{align*}
(-i\omega)a_1 = \left(-i\omega_1 - \frac{\kappa_1}{2}\right)a_1 - i g_1 m - \frac{\kappa_1^{\mathrm{(cpl)}}}{2}a_1 - \frac{\sqrt{\kappa_1^{\mathrm{(cpl)}} \kappa_2^{\mathrm{(cpl)}}}}{2} a_2 +\sqrt{\kappa_1^{\mathrm{(cpl)}}} a^{\mathrm{(in)}}, \\
(-i\omega)a_2 = \left(-i\omega_2 - \frac{\kappa_2}{2}\right)a_2 - i g_2 m - \frac{\kappa_2^{\mathrm{(cpl)}}}{2}a_2 - \frac{\sqrt{\kappa_1^{\mathrm{(cpl)}} \kappa_2^{\mathrm{(cpl)}}}}{2} a_1 +\sqrt{\kappa_2^{\mathrm{(cpl)}}} a^{\mathrm{(in)}}, \\
(-i\omega) m = -i g_1 a_1 - i g_2 a_2 +\left(-i\omega_m - \frac{\gamma_m}{2}\right)m, \\
a_{in}+a_{out} =  \sqrt{\kappa_1^{\mathrm{(cpl)}}} a_1 + \sqrt{\kappa_2^{\mathrm{(cpl)}}} a_2.
\end{align*}
where $\omega_1$, $\omega_2$, and $\omega_m$ are resonant frequencies of the cavity modes and the magnon mode, respectively, $\kappa_1$ and $\kappa_2$ are dissipation rates of the cavity modes, and $\gamma_m$ is the dissipation rate of the magnon mode. 

By solving those equations, we find the following expression for $S_{11}$:

$$
\frac{a_{out}}{a_{in}} = -1+\frac{\kappa_1^{\mathrm{(cpl)}}F_2 + \kappa_2^{\mathrm{(cpl)}}F_1 - 2 \sqrt{\kappa_1^{\mathrm{(cpl)}} \kappa_2^{\mathrm{(cpl)}}}F_3}{F_1 F_2 - F_3^2},
$$
where the following notations are used:
\begin{align*}
F_1 &= i(\omega_1-\omega)+\frac{\kappa_1+\kappa_1^{\mathrm{(cpl)}}}{2} + \frac{g_1^2}{i(\omega_m-\omega)+\frac{\gamma_m}{2}}, \\
F_2 &= i(\omega_2-\omega)+\frac{\kappa_2+\kappa_2^{\mathrm{(cpl)}}}{2} + \frac{g_2^2}{i(\omega_m-\omega)+\frac{\gamma_m}{2}}, \\
F_3 &= \frac{\sqrt{\kappa_1^{\mathrm{(cpl)}} \kappa_2^{\mathrm{(cpl)}}}}{2} + \frac{g_1 g_2}{i(\omega_m-\omega)+\frac{\gamma_m}{2}}.
\end{align*}

\newpage
\section{Temperature dependence of the magnon-photon coupling strength in the ferrimagnetic phase}

\begin{figure*}[!htbp]
\includegraphics[width=1.0\textwidth]{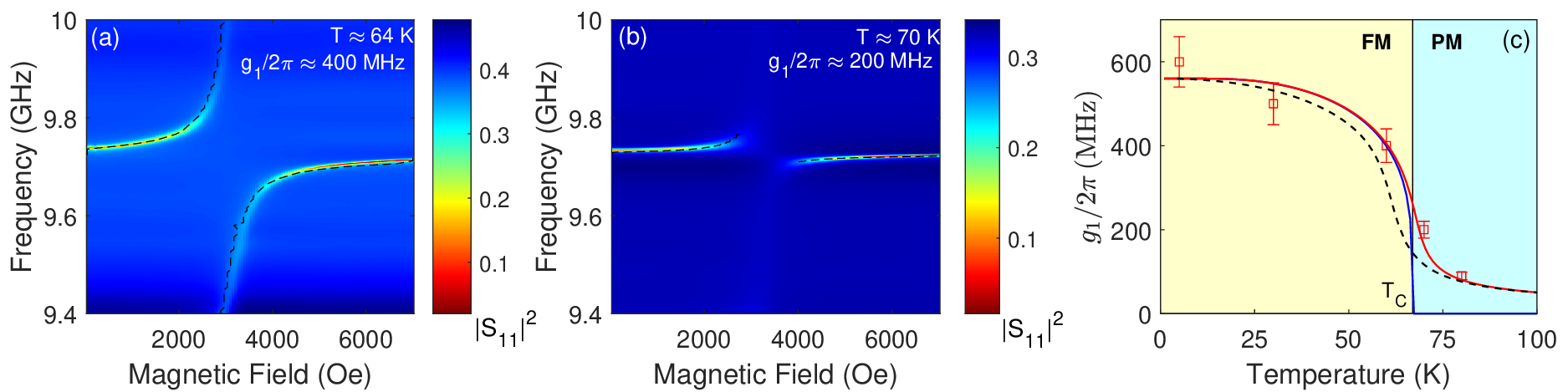}
\caption{\label{figure-g-temperature-dependence}(color online) (a) Microwave response $|S_{11}|^2$ at 64\,K. Dashed lines correspond to theoretical estimates calculated by using equations described in the main text and $g_1/2\pi \approx 400$\,MHz. (b) Microwave response $|S_{11}|^2$ at 70\,K. Dashed lines correspond to theoretical estimates calculated by using the value $g_1/2\pi \approx 200$\,MHz. (c) Temperature dependence of the coupling strength $g_1$ in ferrimagnetic (FM) and paramagnetic (PM) phases: experimental data (squares) obtained in the zero-field-cooled regime and theoretical estimations for $g_1 \propto \sqrt{M(T)}$ for $H=0$\,Oe (solid blue line) and $H=3000$\,Oe (solid red line). The dashed black line corresponds to an estimation of the coupling strength based on the magnetization profile of a Cu$_2$OSeO$_3$ sample of isotropic shape measured at the magnetic field of $H_0=3000$\,Oe, $H_0 \mathrm{\parallel [110]}$.}
\end{figure*}

The coupling strengths $g_1$ and $g_2$ depended strongly on temperature. In particular, the temperature dependence of the coupling strength $g_1(T)$ is shown in Fig.\,\ref{figure-g-temperature-dependence} (measurements were performed in the zero-field-cooled regime). Assuming that the effective total number of spins is proportional to the net magnetization $M$, $N(T) \propto M(T)$, we find that the experimental data can be well described by the $\sqrt{M(T)}$ function normalized to the value of coupling strength at low temperatures (see Fig.\,\ref{figure-g-temperature-dependence}(c)), where $M(T)$ is calculated using the Weiss model of ferromagnetism described below. In our experiments, it was not possible to measure the magnetization profile $M(T)$ directly. However, in Fig. \,\ref{figure-g-temperature-dependence}(c), we plot a theoretical estimate of the temperature dependence of the coupling strength based on the magnetization profile of a Cu$_2$OSeO$_3$ sample of isotropic shape which we measured at the magnetic field of $H_0=3000$\,Oe, $H_0 \mathrm{\parallel [110]}$, using a standard SQUID measurement system. A slight discrepancy between the SQUID data and our results is caused by 1) different demagnetization factors of the two samples, and 2) a difference between the actual temperature of the sample and the value measured by a temperature sensor mounted on the wall of a helium-flow cryostat used in our experiments.
 
\begin{figure}[h]
\includegraphics[width=10cm]{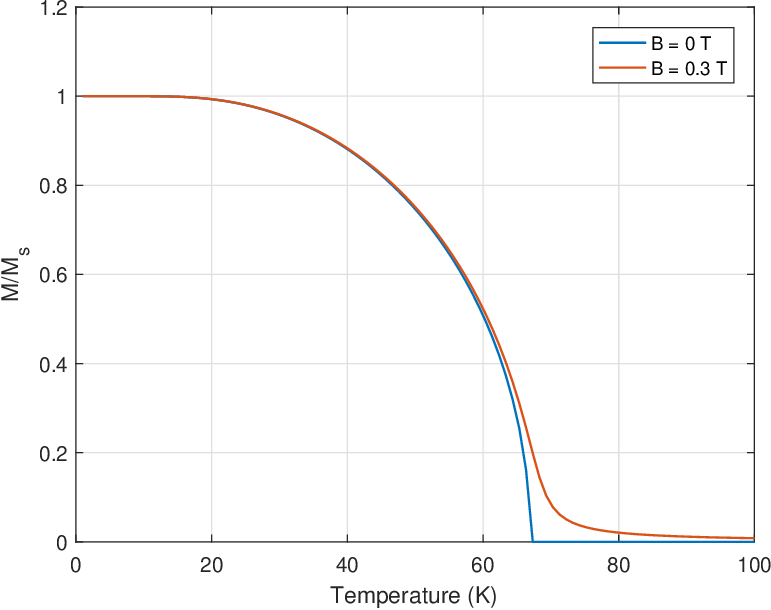}
\caption{\label{figure-magnetization}(color online) Temperature dependence of magnetization at different values of the external magnetic field: 0 T (blue curve) and 0.3 T (red curve) according to numerical calculations.}
\end{figure}

We will use the Weiss model of ferromagnetism\,\cite{Blundell2001} to calculate the net magnetization $M$ in the ferrimagnetic phase as a function of the temperature $T$ and the external magnetic field $B$, $M = M(T,B)$. The ferromagnetic interaction is modeled by introducing the molecular field
$$
B_{mf} = \lambda M,
$$
where $\lambda$ is a constant parameter. For a ferromagnet, $\lambda > 0 $.

The magnetization $M$ can be found by solving self-consistently the following non-linear equations:

\begin{align}
M = M_s B_J(y), \label{MF-model1} \\
y = \frac{g_J \mu_B J (B+\lambda M)}{k_B T}, \label{MF-model2}
\end{align}
where the saturation magnetization $M_s$ is
$$
M_s = n g_J \mu_B J,
$$
$B_J(y)$ is the Brillouin function given by
$$
B_J(y)=\frac{2J+1}{2J}\mathrm{coth}\left(\frac{2J+1}{2J} y \right) - \frac{1}{2J}\mathrm{coth}\left(\frac{y}{2J} \right), 
$$
and $g_J$ is a g-factor, $\mu_B$ is the Bohr magneton, $J$ is a spin number, $k_B$ is the Boltzmann constant, $T$ is the temperature, and $n$ is the spin density.

It can be shown that the parameter $\lambda$ is related to the Curie temperature $T_{\rm C}$ as \cite{Blundell2001}
$$
\lambda = \frac{3 k_B T_{\rm C}}{g_J \mu_B (J+1) M_s}.
$$

For Cu$_2$OSeO$_3$, $J=1$, $g_J\approx2$, $n=4/a^3$ (lattice constant $a=8.93${\AA}), and, by assuming that $T_{\rm C}\approx67$\,K at the magnetic field $B_0 \approx 0.3$\,T, equations\,(\ref{MF-model1}-\ref{MF-model2}) can be solved numerically. Results of numerical calculations for magnetic field values 0\,T and 0.3\,T are shown in Fig.\,\ref{figure-magnetization}.

\newpage
\section{Effect of the magnetic permeability of the material on the cavity resonance frequency}

\begin{figure}[h]
\includegraphics[width=0.9\textwidth]{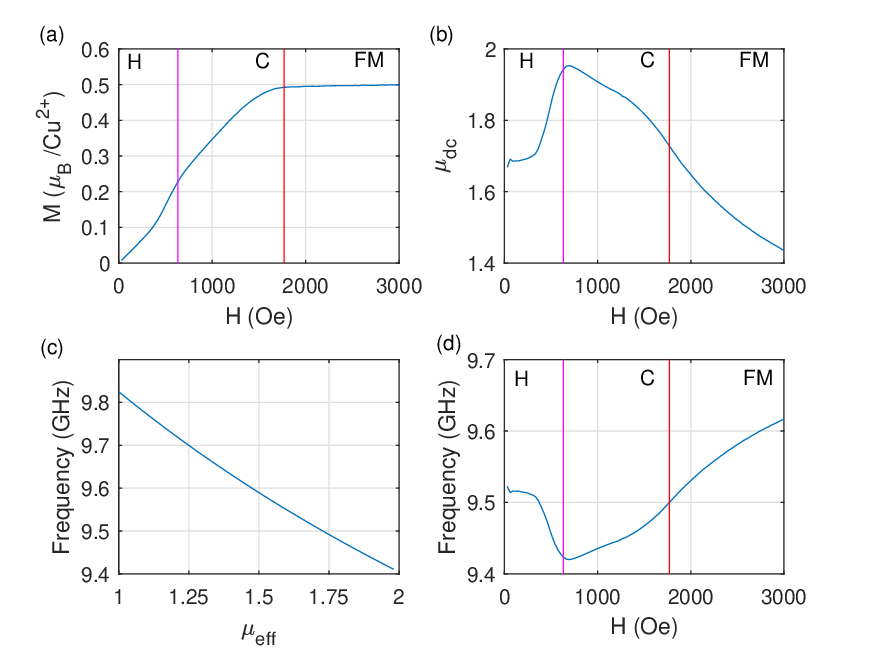}
\caption{\label{figure-permeability}(color online) Simulation of the effect of the dc permeability on the cavity resonance frequency in the case $H \mathrm{\parallel [111]}$. (a) Magnetic field dependence of the magnetization at 5\,K (data is taken from\,\cite{Seki2012}). (b) The dc magnetic permeability $\mu_{dc}$ calculated from the magnetization data. (c) The dependence of the cavity resonance frequency on the effective magnetic permeability $\mu_{eff}$ according to numerical simulations in CST. (d) The calculated dependence of the cavity resonance frequency on the magnetic field caused by the variation of the dc permeability (under assumption $\mu_{eff}=\mu_{dc}$). The calculated data is not consistent with the behavior of the cavity mode observed in our experiments. Thus, experimental data presented in the main text cannot be explained by the effect of the dc permeability. Letter symbols H, C, and FM denote helical, conical, and ferrimagnetic states, respectively. Vertical lines correspond to the transitions between magnetic phases.} 
\end{figure}

In SI units, the relative dc magnetic permeability can be calculated from the field dependence of the net magnetization as $\mu_{dc} = 1+M/H$. In our experiments, the direct measurement of the magnetization $M(H)$ was not possible, therefore in our analysis we will use the data from\,\cite{Seki2012} which was obtained in the experiments with $H \mathrm{\parallel [111]}$. In our measurements, the orientation between the applied magnetic field $H$ and crystal axes was arbitrary, therefore the magnetic fields of the magnetic phase transitions observed in our experiments were different from ones shown in Fig.\,\ref{figure-permeability}. However, the dependence of the dc permeability on the applied magnetic field in our experiments should be qualitatively similar to the one presented in Fig.\,\ref{figure-permeability}(b).

Obviously, there is a significant difference between the experimental data on the magnetic field dependence of cavity resonance frequencies presented in the main text and the one shown in Fig.\,\ref{figure-permeability}(d). In our experiments, the {\it bare} cavity mode frequency did not depend on the magnetic field --- values of cavity resonance frequencies at very low magnetic fields and at very high magnetic fields were very close to each other. Therefore, we can make a conclusion that the cavity mode was not affected by the dc permeability.

However, it should be emphasized that microwave response of magnetic materials can be described using an ac permeability\,\cite{Queffelec2002,Krupka2016}, and many aspects of magnon-photon coupling phenomena can be explained in terms of an effective ac permeability\,\cite{Hyde2017}.

%

\widetext
\clearpage
\setcounter{equation}{0}
\setcounter{figure}{0}
\setcounter{table}{0}
\setcounter{page}{1}
\makeatletter
\renewcommand{\theequation}{S2-\arabic{equation}}
\renewcommand{\thefigure}{S2-\arabic{figure}}
\renewcommand{\bibnumfmt}[1]{[S2-#1]}
\renewcommand{\citenumfont}[1]{S2-#1}

\begin{center}
\textbf{\large Supplemental Material for ``Magnon-photon coupling in a non-collinear magnetic insulator Cu$_2$OSeO$_3$''. Part 2: Calculations of the overlap integral $\eta_m$.}

Masahito Mochizuki

\textit{Department of Applied Physics, Waseda University, Okubo, Shinjuku-ku, Tokyo 169-8555, Japan}

Numerical calculations of the overlap integrals $\eta_m$ for magnon modes in Cu$_2$OSeO$_3$.
\end{center}

\section{Magnetism of the chiral cubic Copper Oxoselenite}
A spin model for Cu$_2$OSeO$_3$ is constructed as follows. The crystal structure of this compound belongs to the chiral cubic P2$_1$3 space group, while the magnetic structure is comprised of tetrahedra composed of four Cu$^{2+}$ ($S$=1/2) ions at their apexes. The Cu spins in each tetrahedron form a three-up and one-down ferrimagnetic ($S$=1) configuration below $T_{\rm c}\sim$58 K, and this spin assembly can be regarded as a magnetic unit because the intra-tetrahedron spin couplings are much stronger than the inter-tetrahedron spin couplings. The crystallographic unit cell of this compound is a cube whose volume is $a^3$ with the lattice constant of $a$=8.93 \AA. Because the unit cell contains four tetrahedra, each tetrahedron occupies a spatial volume of $a_0^3$ where $a_0$=$a/r$ with $r=4^{1/3}$.

\section{Spin Model and coarse graining}
The magnetism of Cu$_2$OSeO$_3$ is described by a classical Heisenberg model on a cubic lattice. The Hamiltonian is given by,
\begin{eqnarray}
\mathcal{H}_0
&=&-J_0 \sum_{i,\bm \gamma=\hat{\bm x}, \hat{\bm y}, \hat{\bm z}}
\bm m_i \cdot \bm m_{i+\hat{\bm \gamma}}
\nonumber \\
& &-D_0 \sum_{i,\bm \gamma=\hat{\bm x}, \hat{\bm y}, \hat{\bm z}}
(\bm m_i \times \bm m_{i+\hat{\bm \gamma}}) \cdot \hat{\bm \gamma}
\nonumber \\
& &-g\mu_{\rm B}S \bm H_0 \cdot \sum_i \bm m_i
\nonumber \\
& &+K_0 \sum_i \left(m_{ia}^4+m_{ib}^4+m_{ic}^4 \right)
\label{eq:smd1}
\end{eqnarray}
where $\bm m_i$ is a normalized classical magnetization vector for the $i$th tetrahedron. The first term describes the ferromagnetic exchange interactions, while the second term depicts the Dzyaloshinskii-Moriya interactions with $\hat{\bm x}$, $\hat{\bm y}$, and $\hat{\bm z}$ being the directional vectors along the $x$, $y$ and $z$ axes. The third term represents the Zeeman interactions associated with an external magnetic field $\bm H_0$ where $g(=2)$ is the Lande's g-factor, $\mu_0$ is the magnetic permeability for vacuum, and $S(=1)$ is the spin amplitude per tetrahedron. The last term denotes the fourth-order magnetic anisotropy allowed in the cubic crystals where $a$, $b$ and $c$ are the orthogonal coordinates in the cubic setting. 

For slowly varying spin textures in the helical, conical and skyrmion phases, the spins are nearly decoupled from the background lattice structure. It justifies our theoretical treatment based on a spin model on the cubic lattice for simplicity without considering complicated crystal structures of real materials. This spin Hamiltonian exhibits a phase transition between the paramagnetic and the helical states at $k_{\rm B}T_{\rm c}$=1.43$J_0$ when $\bm H_0$=0. From the experimentally observed $T_{\rm c}\sim$58 K, $J_0$ is evaluated as $J_0$=3.50 meV.

As far as the slowly varying spin textures are concerned, this lattice spin model is rewritten in a continuum form as,
\begin{eqnarray}
\mathcal{H}_1&=&\int d^3\bm r\; 
 \frac{J_0}{2a_0}(\bm \nabla m)^2
+\frac{D_0}{a_0^2}\bm m \cdot (\bm \nabla \times \bm m)
\nonumber \\
& &-\frac{g\mu_{\rm B}S}{a_0^3} \bm H_0 \cdot \bm m
+\frac{K_0}{a_0^3}\sum_i \left(m_a^4+m_b^4+m_c^4 \right)
\end{eqnarray}
This expression indicates that we can reconstruct a lattice spin model by regarding an assembly of magnetizations in a cubic cell with volume of $a^3$=$(ra_0)^3$ (instead of a single spin or a single magnetic unit) as one magnetization. This coarse graining again gives a classical Heisenberg model on a cubic lattice as,
\begin{eqnarray}
\mathcal{H}_{\rm c.g.}
&=&-J \sum_{i,\bm \gamma=\hat{\bm x}, \hat{\bm y}, \hat{\bm z}}
\bm m_i \cdot \bm m_{i+\hat{\bm \gamma}}
\nonumber \\
& &-D \sum_{i,\bm \gamma=\hat{\bm x}, \hat{\bm y}, \hat{\bm z}}
(\bm m_i \times \bm m_{i+\hat{\bm \gamma}}) \cdot \hat{\bm \gamma}
\nonumber \\
& &-g\mu_{\rm B}S \bm H \cdot \sum_i \bm m_i
\nonumber \\
& &+K \sum_i \left(m_{ia}^4+m_{ib}^4+m_{ic}^4 \right).
\end{eqnarray}
Here $\bm m_i$ is a normalized classical vector describing a representative magnetization in the $i$th cubic cell. We find that following relationships hold between coupling constants of the original spin model $\mathcal{H}_1$ and those of the coarse-grained spin model $\mathcal{H}_{\rm c.g.}$,
\begin{eqnarray}
J=rJ_0, \; D=r^2D_0, \; \bm H=r^3 \bm H_0, \; K=r^3 K_0.
\end{eqnarray}
When we adopt the crystallographic unit cell ($a$=8.93 \AA) as a cubic cell for the coarse graining, $r=a/a_0$ becomes $4^{1/3}$ because the unit cell contains four tetrahedra.

In the following calculations, we adopt $J$=$rJ_0$=5.56 meV as the energy units, and we use $D/J$=-0.09 and $K/J$=-0.003. This Dzyaloshinskii-Moriya parameter reproduces the experimentally observed helical periodicity of $\lambda_{\rm H}$=63 nm. The ground-state energy of the helical order is given as a function of the spin rotation angle $\theta$ as
\begin{eqnarray}
E(\theta)/N=-J\cos\theta-D\sin\theta.
\end{eqnarray}
Solving the saddle-point equation
\begin{eqnarray}
\frac{dE(\theta)}{d\theta}=J\sin\theta-D\cos\theta=0,
\end{eqnarray}
we obtain a relationship $\tan\theta$=$D/J$ for the helical and conical states with a minimum energy, which gives $\theta$=5.142$^\circ$ and $\lambda_{\rm H}$=$(360^\circ/\theta)a$=63 nm when $D/J$=0.09 and $a$=8.93 \AA. This set of parameters also reproduces the experimentally observed critical magnetic field of $\sim$750 Oe for the phase transition between the conical and the ferrimagnetic states at low temperatures. In the numerical simulations, we found that the phase transition occurs at $g\mu_{\rm B}SH_c$=$0.006J$, which corresponds to $H_{0c}$=$H_c/r^3$=720 Oe. 
The unit conversions when $J_0$=3.50 meV ($J$=$rJ_0$=5.56 meV) and $S$=1 are summarized in Table~\ref{tab:uconv}.
\begin{table}
\begin{tabular}{l|cc} \hline \hline
Temperature    & $k_{\rm B}T/J=1$  & $T$=$J_0/k_{\rm B}$=40.6 K \\
Magnetic field & $g\mu_{\rm B}H/J=0.001$ & $H_0$=$H/r^3$=120 Oe \\
Frequency $f=\omega/2\pi$ & $\hbar\omega/J=0.001$ & $f$=1.35 GHz\\
Time & $tJ/\hbar=1$ & $t$=0.12 ps\\
\hline \hline
\end{tabular}
\caption{Unit conversion table when $J$=$rJ_0$=5.56 meV and $S$=1.}
\label{tab:uconv}
\end{table}

We define the $x$, $y$ and $z$ axes oriented as $\bm x \parallel [1\bar{2}1]$, $\bm y \parallel [10\bar{1}]$, and $\bm z \parallel [111]$. With these orthogonal coordinates, the fourth-order magnetic anisotropy term is rewritten in the form,
\begin{eqnarray}
K\sum_i \left(
\frac{1}{2}+m_{iz}^2-\frac{7}{6}m_{iz}^4-\frac{2\sqrt{2}}{3}m_{iz}
(m_{ix}^3-3m_{ix}m_{iy}^2)
\right)
\end{eqnarray}
For the external magnetic field $\bm H$, we consider both the static magnetic field $\bm H_{\rm ext}=(0, 0, H_z)$ applied along the $z$ axis and the time-dependent magnetic field $\bm H(t)$.

\section{Results for resonance modes}
We calculate dynamical magnetic susceptibilities to evaluate the resonance frequencies of magnon modes. We first performed Monte-Carlo thermalizations for the Hamiltonian $\mathcal{H}_{\rm c.g.}$ to obtain stable magnetization configurations at low temperatures ($k_{\rm B}T/J$=0.01) for various values of $H_z$. Then we further relaxed them in the micromagnetic simulations using the Landau-Lifshitz-Gilbert (LLG) equation to obtain the ground-state magnetization configurations. The LLG equation was solved by the fourth-order Runge-Kutta method. A system of 20$\times$20$\times$140 sites with periodic boundary conditions were used for the numerical calculations. The LLG equation is given by,
\begin{equation}
\frac{d\bm m_i}{dt}=-\gamma \bm m_i \times \bm H^{\rm eff}_i 
+\frac{\alpha_{\rm G}}{m} \bm m_i \times \frac{d\bm m_i}{dt}.
\label{eq:LLGEQ}
\end{equation} 
Here $\alpha_{\rm G}$ is the Gilbert-damping coefficient. The effective magnetic field $\bm H_i^{\rm eff}$ acting on the local magnetization $\bm m_i$ on the $i$th cell is calculated from the Hamiltonian $\mathcal{H}_{\rm c.g.}$ in the form
\begin{equation}
\bm H^{\rm eff}_i = -\frac{1}{\gamma\hbar r^3}\frac{\partial \mathcal{H}_{\rm c.g.}}{\partial \bm m_i}.
\label{eq:EFFMF}
\end{equation}

The dynamical magnetic susceptibilities for a microwave field $\bm H(t) \parallel \bm y$ perpendicular to the external static magnetic field $\bm H_{\rm ext}$ is given by,
\begin{eqnarray}
\chi_y(\omega) =
\frac{\Delta M_y(\omega)}{H_y(\omega)}
\end{eqnarray}
Here $H_y(\omega)$ and $\Delta M_y(\omega)$ are Fourier transforms of the time-dependent magnetic field $\bm H(t)$ and the simulated time-profile of the total magnetization $\Delta \bm M(t)=\bm M(t)-\bm M(0)$ with $\bm M(t)=\frac{1}{N}\sum_{i=1}^N \bm m_i(t)$. For these calculations, we use a short rectangular pulse for the time-dependent field $\bm H(t)$ whose components are given by,
\begin{eqnarray}
H_y(t)=\left\{
\begin{aligned}
& H_{\rm pulse} \quad & 0 \le tJ/\hbar \le 1 \\
& 0 \quad & {\rm others}
\end{aligned}
\right.
\end{eqnarray}
An advantage of using the short pulse is that for a sufficiently short duration $\Delta t$ with $\omega \Delta t \ll 1$, the Fourier component $H_y(\omega)$ becomes constant being independent of $\omega$ up to the first order of $\omega \Delta t$. The Fourier component is calculated as 
\begin{eqnarray}
H_y(\omega)&=&\int_0^{\Delta t}\;H_{\rm pulse} e^{i\omega t}dt
=\frac{H_{\rm pulse}}{i\omega}\left(e^{i\omega \Delta t}-1 \right)
\nonumber \\
&\sim&H_{\rm pulse} \Delta t.
\end{eqnarray}
As a result, we obtain the relationship $\chi_y(\omega) \propto \Delta M_y(\omega)$.

\begin{figure*}
\includegraphics[scale=1.0]{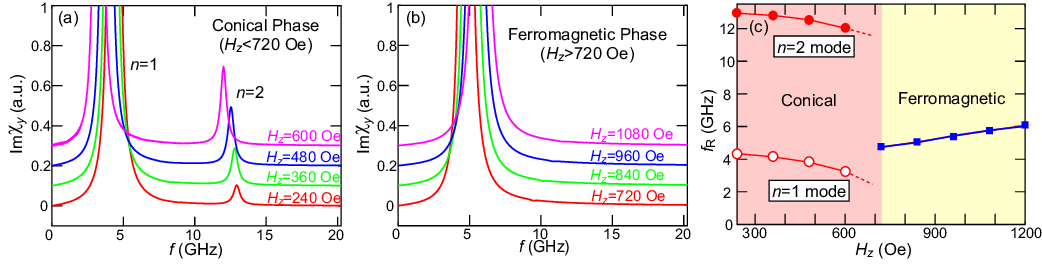}
\caption{(a), (b) Calculated imaginary parts of the dynamical magnetic susceptibilities in (a) the conical phase and (b) the ferromagnetic phase for various magnetic-field strengths $H_z$. For the conical phase, the resonance peaks of the $n$=2 mode appear in addition to the large peaks of the lower-lying $n$=1 mode. (c) Resonance frequencies of the conical $n$=1 and $n$=2 modes and the ferromagnetic resonance mode as functions of $H_z$.}
\label{Fig01MM}
\end{figure*}
In Fig.~\ref{Fig01MM}(a) and (b), we show calculated imaginary parts of the dynamical magnetic susceptibilities Im$\chi_y$ for the conical phase and the ferromagnetic phase respectively. We find that in the conical phase, a tiny peak corresponding to the $n$=2 mode appears in each spectrum in addition to the large peak of the lower-lying $n$=1 mode. In Fig.~\ref{Fig01MM}(c) we plot the calculated resonance frequencies as functions of $H_z$.

\section{Results for the overlap integrals}
Following works\,\cite{Lambert2015,Zhang2016}, we calculate the overlap integral $\eta$ which describes the extent of an overlap between a microwave and a magnon mode. This quantity is given by,
\begin{eqnarray}
\eta(t)=
\left|
\frac{\int\; \bm H(\bm r, t) \cdot \Delta \bm m(\bm r, t) d\bm r}
{{\rm max}|\bm H(\bm r, t)|\;{\rm max}|\Delta m_{\parallel}(\bm r, t)|V}
\right|,
\label{eta}
\end{eqnarray}
where $\bm H(\bm r, t)$ is a microwave magnetic field, $\Delta \bm m(\bm r, t)\equiv\bm m(\bm r, t)-\bm m(\bm r, 0)$ is an oscillating component of the local magnetization $\bm m(\bm r, t)$, $\Delta m_{\parallel}(\bm r, t)$ is a component of $\Delta \bm m$ parallel to $\bm H(\bm r, t)$, and $V$ is the system volume. Here ${\rm max}|\bm H(\bm r, t)|$ and ${\rm max}|\Delta m_{\parallel}(\bm r, t)|$ denote maximum values in the spatial volume at each time $t$.
In the case of continuous excitations, this quantity converges at a constant value and becomes time independent when the excitation becomes steady after a sufficiently long duration. 

\begin{figure*}
\includegraphics[scale=1.0]{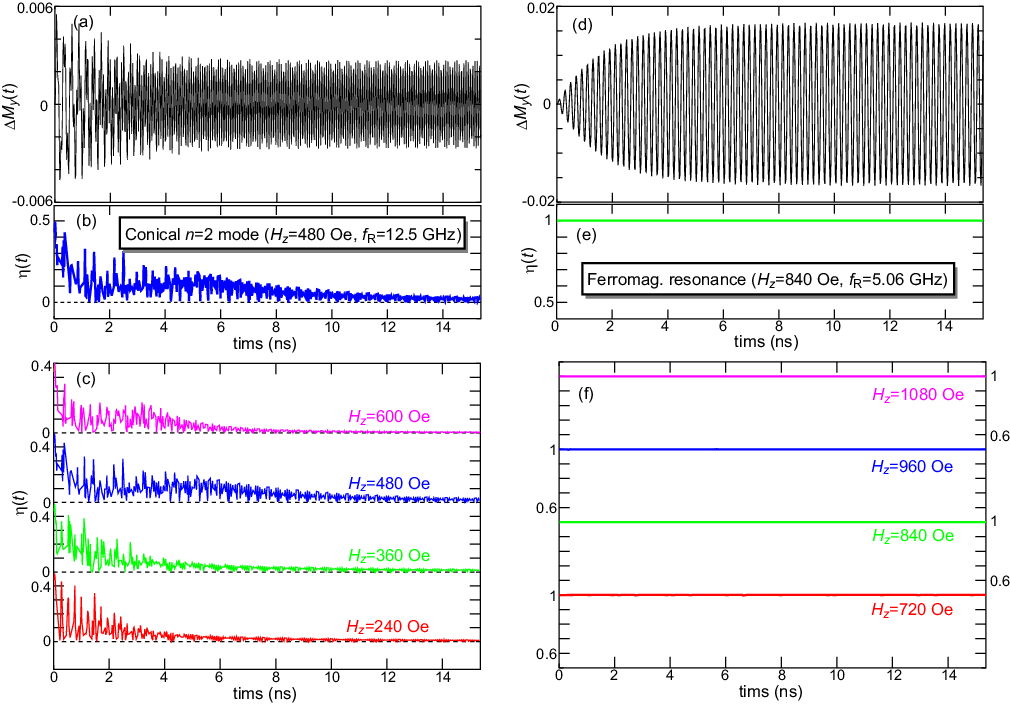}
\caption{(a), (d) Simulated time profiles of oscillating component of the net magnetization $\Delta \bm M(t)=\bm M(t)-\bm M(0)$ with $\bm M(t)=(1/N)\sum_i \bm m_i(t)$ under application of the $y$-polarized microwave field with a resonant frequency for (a) the conical $n$=2 mode when $H_z$=480 Oe and (b) the ferromagnetic resonance mode when $H_z$=840 Oe. (b), (e) Calculated time evolutions of the microwave-magnon overlap integral $\eta$ for (b) the conical $n$=2 mode when $H_z$=480 Oe and (e) the ferromagnetic resonance mode when $H_z=840$ Oe. (c), (f) Time evolutions of $\eta$ for various magnetic-field strengths $H_z$. Those for the conical $n$=2 mode always converge at a very tiny value of $\sim$0.01 after a sufficient duration, whereas those for the ferromagnetic resonance are constantly unity.}
\label{Fig02MM}
\end{figure*}
To calculate this quantity, we simulate time profiles of $\Delta \bm m(\bm r, t)$ under application of a sinusoidally oscillating microwave field $\bm H(t) \parallel \bm y$ by numerically solving the Landau-Lifshitz-Gilbert equation using the fourth-order Runge-Kutta method. The $y$-axis component of $\bm H(t)$ is given by
\begin{eqnarray}
H_y(t)=H^\omega \sin \omega t.
\end{eqnarray}
Here the amplitude of microwave field is fixed at $g\mu_{\rm B}SH^\omega/J=0.00001$ when calculating $\eta$ for the conical $n$=1 mode and the ferromagnetic resonance mode, whereas at $g\mu_{\rm B}SH^\omega/J=0.0002$ for the conical $n$=2 mode. The Gilbert-damping coefficient is fixed at $\alpha_{\rm G}=0.02$. We performed the calculation for various resonance modes and magnetic-field strengths $H_z$ by substituting the corresponding resonance frequencies $\omega_{\rm R}$ into $\omega$. 

In Fig.~\ref{Fig02MM}(a) and (d), we show simulated time profiles of oscillating component of the net magnetization $\Delta \bm M(t)=\bm M(t)-\bm M(0)$ with $\bm M(t)=(1/N)\sum_i \bm m_i(t)$ under application of the $y$-polarized microwave field with a resonant frequency for the conical $n$=2 mode when $H_z$=480 Oe and the ferromagnetic resonance mode when $H_z$=840 Oe, respectively. We find that steady oscillations occur after a duration of $\sim$4 ns. Accordingly, the time evolution of $\eta$ for the conical $n$=2 mode in Fig.~\ref{Fig02MM}(b) is time-dependent, whereas that for the ferromagentic resonance mode in Fig.~\ref{Fig02MM}(e) takes constantly unity. These mode-specific behaviors do not alter even when the magnetic-field strength $H_z$ varies as seen in Fig.~\ref{Fig02MM}(c) and (f) which summarize the simulated time profiles of $\eta$ for various values of $H_z$.

\begin{figure}
\includegraphics[scale=1.0]{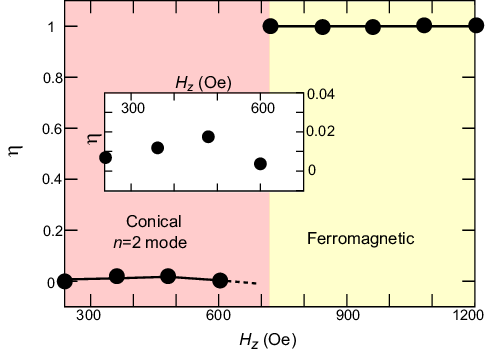}
\caption{Calculated values of microwave-magnon overlap integrals $\eta(t)$ at the moment of time $t$=15.36 ns for the conical n=2 mode and the ferromagnetic resonance mode as functions of the magnetic-field strength $H_z$. The inset magnifies the plots for the conical $n$=2 mode.}
\label{Fig03MM}
\end{figure}

Although it is difficult to evaluate the precise saturation value of $\eta(t)$ by the exponential fitting, the time profiles of $\eta(t)$ shown in Fig.~\ref{Fig02MM}(c) indicate that the mode overlapping for the n=2 helimagnon mode is significantly suppressed as compared to the one for the ferromagnetic resonance mode. However, the value is still expected to be finite as evidenced by the time profile of $\Delta M_y(t)$ in Fig.~\ref{Fig02MM}(a), which shows a finite amplitude of the steadily oscillating $n$=2 mode, indicating the finite coupling between the microwave driving field and the $n$=2 helimagnon mode. We expect that $\eta$ for the $n$=2 helimagnon mode is typically of the order of 0.01 according to the plot in the Fig.~\ref{Fig03MM} which shows the field-dependence of $\eta(t)$ after sufficient duration ($t$=15.56 ns).

\end{document}